# IMPACTS OF CULTURE AND SOCIO-ECONOMIC CIRCUMSTANCES ON USERS' BEHAVIOR AND MOBILE BROADBAND TECHNOLOGY DIFFUSIONS TRENDS
# A COMPARISON BETWEEN THE UNITED KINGDOM (UK) AND BANGLADESH

Mahdi H Miraz, Monir Bhuiyan and Md. Emran Hossain

Department of Computer Science & Software Engineering, University of Hail, KSA
`m.miraz@uoh.edu.sa`
Institute of Information Technology, University of Dhaka, Bangladesh
`mbhuiyan@iit.du.ac.bd`
Research Centre for Technologies in Society,CICTR,Bangladesh
`emranni@yahoo.com`

## ABSTRACT

*The use of Internet and Internet-based services on PCs, Laptops, Net Pads, Mobile Phones, PDAs etc have not only changed the global economy but also the way people communicate and their life styles. It also has evolved people from different origins, cultures, beliefs across the national boundaries. As a result it has become an absolute necessity to address the cross-cultural issues of information systems (IS) reflecting the user behaviours and influencing the way the mobile broadband technology is being accepted as well as the way it is changing the life styles of different groups of people. This paper reports on an on-going research effort which studies the impacts of culture and socio-economic circumstances on users' behavior and mobile broadband technology diffusions trends.*

## 1. INTRODUCTION

The web has become a commodity that everyone has to have and everyone needs to use because it's built upon the most important commodity of the next millennium: information [1]. With the phase of time, People are also very quickly moving towards mobile computing. Not only it lets us use the internet on the move by adding more flexibility but also some applications are based on mobile internet without which life cannot be imagined today. The increasing demand for mobile broadband access to multimedia and Internet applications and services over the last few years has created new interest among existing and emerging operators to explore new technologies and network architectures to offer such services at low cost to operators and end users. [2]. It has now became extremely difficult in 21$^{st}$ century to ignore to address the Information System (IS) issues relating to cultural differences, communication barriers, different Human Computer Interaction (HCI) principles along with user behaviour, socio-economic circumstances etc . The purpose of this research is to have an in depth look into these issues and the paper is going to present a brief discussion along with some suggestions to overcome these issues.

This paper is only going to focus on how cultural and socio-economic circumstances are being reflected on the behaviour of IS users across different national boundaries and the diffusion of mobile broadband technology (including internet-based services) due to this. We have focused our case among the IS users of the United Kingdom and Bangladesh.

## 2. RESEARCH METHODOLOGY

A qualitative approach was used to study the socio-economical and cross-cultural IS issues of the focused group. The research was carried out by direct observations of a focus group, conducted during one-to-one tutoring sessions, structured and unstructured interviews were also included. The multiple case-study approach was adopted to increase the reliability of data, and a team of researchers and tutors were employed to reduce bias [3][4]. In this case study, total 123 IS users had participated. Among them, 57 Bangladeshi and 53 UK IS users participated into the survey. 13 Bangladeshi IT professionals also took part in the interview.

We conducted the survey at the end of the study to verify our results. The survey was not only conducted by distributing printed questionnaire but also web based version was used. All the participants were adults and IS users. The questionnaire was designed in two different languages- English and Bangla.

Our sample is representative because the participants are the randomly selected individual internet users from Bangladesh & UK and Bangladeshi IT professionals. Respondents for interviews were selected by using the concept of theoretical sampling; that is, respondents (Bangladeshi IT professionals) are sampled on the basis that they are knowledgeable on the detailed aspects of mobile broadband technologies and also familiar with the cultural and scio-economical situations in Bangladesh. A technology is designed to be used by the users. So Bangladeshi users have been chosen because their feedback will represent the true picture of how they use the technology and what they expect in the future. The reason for choosing UK users is to compare the feedback found from Bangladeshi users. As UK represents a Developed country, the comparison will let us see the difference in user behaviour and technology diffusion.

## 3. BACKGROUND

Bangladesh is a small country located in South Asia. Although, mobile phone operators have reached the very last mile of the country covering in 95% of the total land area, other telecommunication facilities have not yet been equally deployed in the whole country. Country's new ICT policy is aimed to remove these gaps by deploying WiMAX and 3G over the entire country over next few years.

By finally joining the SEA-ME-WE-4 submarine cable network consortium in 2006, Bangladesh has joined rest of the Internet world with a data transfer capacity of 10 gigabytes per second which is definitely much better compared to the 150 megabytes per second bandwidth previously used by the government owned Bangladesh Telegraph and Telephone Board (BTTB) and other Inter Service Providers (ISPs).

Now the big challenge is to improve the country's internal telecommunication infrastructure providing Internet access and the benefits of Internet services to each and every citizens. As the wired infrastructure Bangladesh has got now is not sufficient to fulfil this challenge, mobile or Wireless Broadband Access (WBA) could be a better solution. There are two major reasons for considering mobile or wireless broadband as a probable alternative. Firstly, Setting up mobile or WBA uses air as medium [7], WBA networks do not require polling cables as much as it is required for traditional wired networks. Less wiring obviously requires less money to invest and reduce the amount of hassle needed to setup, run and maintain network in a wired fashion. Secondly, mobile or WBA can reach the 'last mile' covering more areas. This can specially help spreading Internet access in the rural, hilly and river dense areas of the country. Hence, WiMAX and 3G now hold huge importance in the field of Mobile Internet and Wireless Communications of Bangladesh.

Developing countries are far behind in comparison with the developed countries in respect of deploying technologies. The study also focuses on how these technologies can help the developing countries over the traditional and existing infrastructure they have already got. To achieve this goal, Surveys have been conducted to find out the technology adoption trend among Bangladeshi resident. Another similar survey was conducted among the UK residents. The comparison between Bangladeshi and UK resident Internet users let us see how socio-economic circumstances reflect on the users' behaviour and technology diffusion.

## 4. RESULTS AND DISCUSSIONS

The aim of the survey was to find out adoption trend of mobile broadband technologies and demographic information of the internet users in Bangladesh and compare with that of UK. Also opinion and suggestions from IT professionals were sought through 10 scale closed questions. The reasoned justification for using a ten continuous scale is due to the fact that a continuous scale produces more consistent results compared to other scaling. Following are the analysis of the information gathered from the survey along with information from other sources.

### Number of Home Internet Users

The number of people using the internet at home in Bangladesh is very low in comparison with that of United Kingdom. Only 7.0% people (who do not use the internet at work) use the internet at home in Bangladesh. The number in UK is much higher (69.8%) for the same type of users. Our research suggests that the reason of this low number of home users in Bangladesh is the higher price of internet and poor ICT infrastructure.

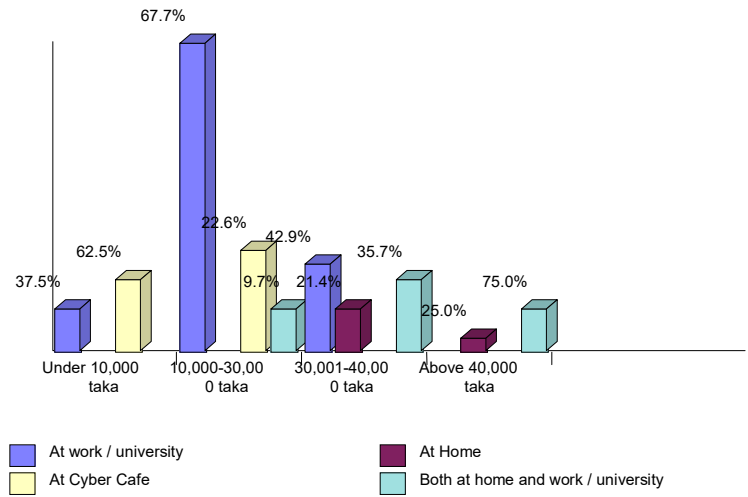

Figure 1: Income vs. Location of Use (Bangladesh)

The survey conducted also indicates that the economic incapacity is one of the main reasons in this case. Figure-1 shows that the highest percentage (21.4% of the total internet users of that earning band) of home users are earning 30K to 40K taka (currency of Bangladesh) per month. It is to mention 1 Pound Sterling (GBP) is equivalent to approximately 115 taka. Those having less family income, either use the internet at work/university or occasionally at any cyber cafe.

**Number of PC users**

Number of people using computer in developing country is much less than that of a developed country. As a result mobile phone sometimes serves as the computer. It has been found that among Bangladeshi users email is the most popular feature used in a mobile phone in comparison with SMS, MMS, and others. 33.3% of the mobile users' favourite feature is Email and 24.6% people's favourite feature is SMS. But in UK, 52.8% people uses SMS as their favourite mobile phone feature and only 9.4% people consider email to be the favourite feature of a mobile phone. This statistics clearly indicates that because people in UK has more access to the computer, they do not need to use mobile phone to use for email services unless otherwise away from home/work or other reasons. But as in Bangladesh, most of the families do not have a PC or even they do have a PC it is unlikely that the PC is connected to the internet. As a result, people tend to use mobile phone as a medium of email service.

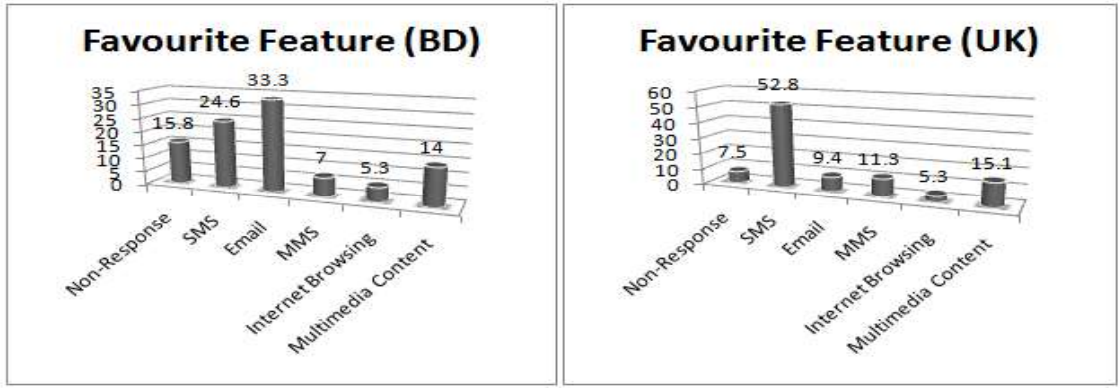

Figure 2: Favourite Feature on Mobile Phone

The Age vs. Favourite Feature also proves this trend. As for instance, only 44.4% of the UK internet users aged between 36 to 45 considers email to be their favourite but in Bangladesh the number is as high as 66.7%. This is because in Bangladesh, these types of internet users are either business persons or professionals and they need to use the internet. Having no other alternatives, they prefer to use the mobile phone.

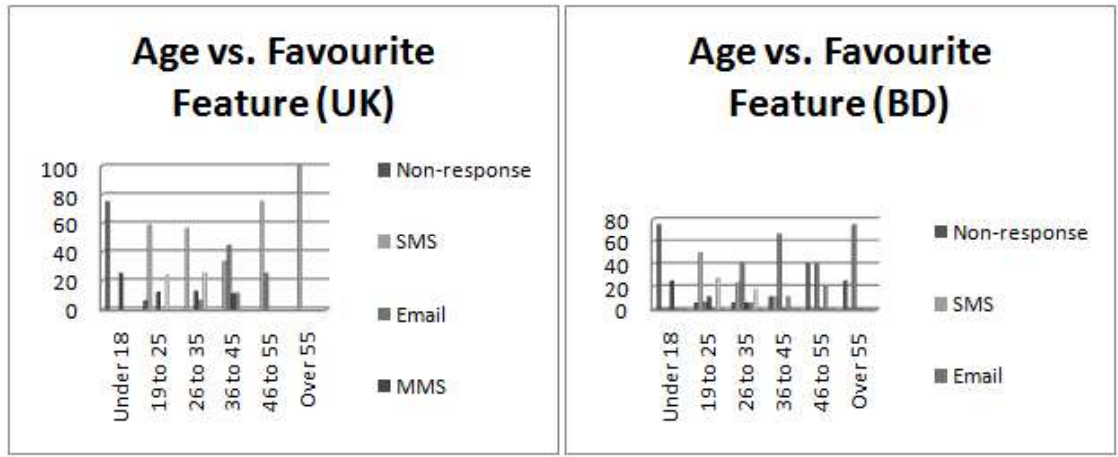

Figure 3: Age vs. Favorite Feature

**Ownership of 3G Phone**

Only 10.5% of the internet users have 3G mobile phone in Bangladesh whereas 35.8% UK resident internet users have a 3G phone. Low number of people having 3G phone and its being expensive will also have a great impact on the diffusion of 3G technologies if it is deployed. Although unfortunately the rate of using WAP-based services is much lower in comparison of the ownership of 3G or WAP phones. One of the UK mobile operators reported that fewer than 5% of WAP phone owners actually use their internet-based services.[9]

| Do you own a 3G phone? | Percentage of Participants (Bangladesh) | Percentage of Participants (UK) |
|---|---|---|
| Yes | 10.5% | 35.8% |
| No | 71.9% | 45.3% |
| Do not Know | 17.6% | 18.9% |

Table 1: Ownership of 3G Phone (Bangladesh)

**Female Education and ICT literacy**

South Asia is one of the few areas in the world where gender discrimination is so severe that aggregate population statistics reveal skewed gender ratios suggesting differential life expectancies between women and men resulting from social, economic and cultural factors [5]. Our study also suggests that there exists a huge "gender divide" among the IS users of Bangladesh. This number in Bangladesh is much low because of low female education rate.[6] Number of female participant is 29.8% (17 out of total 57) in Bangladesh and 39.6% (21 out of total 53) in UK which indicates that because the society in Bangladesh is male dominant, it is also impacting the ICT literacy and technology adoption.

| Sex | Percentage of Participants (Bangladesh) | Percentage of Participants (UK) |
|---|---|---|
| Male | 70.2% | 60.4% |
| Female | 29.8% | 39.6% |

Table 2: Female Participant

**Impacts of Age**

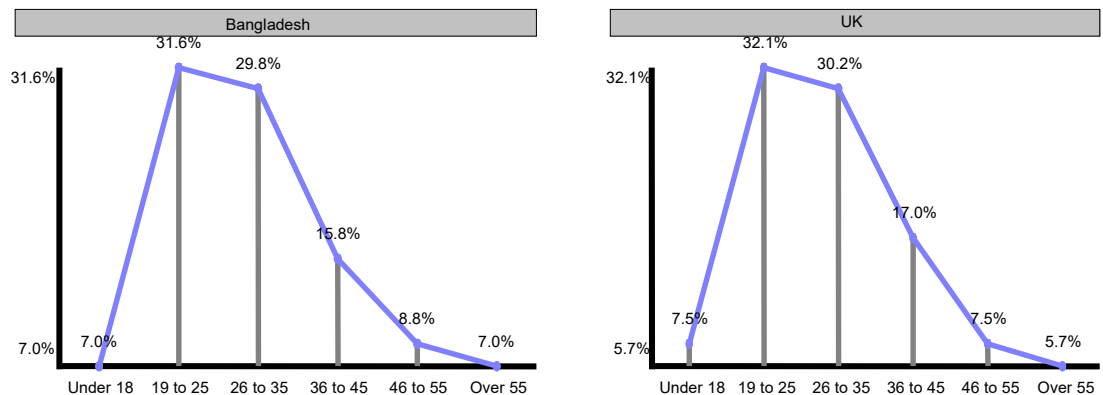

Figure 4: Age vs. Adoption of New Technology

The number of users also depends on age. Although interestingly enough, the study came out with the result that age based trend is similar in both of the countries. The interest for adoption of new technology grows as the user grow up and at a certain stage of age the adoption of new technology starts declining. The graphs in Figure-4 represent the almost similar curve for both the countries.

**Price of the Service vs. Individual's Economic Capacity**

The urban and rural wealthy are likely to have access and the material assets to adopt the state of the art technology.[8] Only 7.6% of the total Bangladeshi internet users fall into this category and have Mobile Broadband Card or Dongle and the same number is about 3 times higher in UK. More than 22.22% people of the total internet users have Mobile broadband card or USB dongle. This is because the price of these services is high and income capacity of Bangladeshi people is low in comparison with that of UK.

| How do you use Mobile Internet? | Percentage of Participants (Bangladesh) | Percentage of Participants (UK) |
|---|---|---|
| On Mobile Phone | 71.79% | 69.44% |
| Using Mobile phone as a | 20.51% | 8.33% |
| Using Mobile Broadband | 7.6% | 22.22% |

Table 3: Method of Using Mobile Internet

43.47% people consider purchasing a 3G phone only with the view of using mobile internet will be too much spending for the service while 32.6% of them cannot afford at all. In UK the number is very low. Only 23.07% people consider it will be too much spending and 19.23% considers they cannot afford it.

| Will you consider purchasing a 3G phone ONLY for the purpose of | Percentage of Participants (Bangladesh) | Percentage of Participants (UK) |
|---|---|---|
| Very Likely | 6.52% | 11.53% |
| Likely | 17.39% | 46.15% |
| Not Likely (Too expensive) | 43.47% | 23.07% |
| Not Likely (Not affordable) | 32.60 | 19.23 |

Table 4: Trend of Purchasing 3G Phone

**IT Professionals' View**

IT professionals consider that large scale deployment of mobile broadband will change the way people use internet now especially for Business, Students and Research purpose with a mean value of 7.47, 8.68 and 9.11 accordingly. Results obtained from the feedback provided on 10 scale form. But they do not consider that it will not change to a great extend the way people use the internet for online shopping, banking or ticketing etc. The mean value obtained for these types of uses are 2.68, 3.11 and 2.79 respectively. The research also found that if mobile broadband is deployed to a large scale, people will be more using the social networking applications (with a mean value of 6.37) like face book, msn, twitter etc.

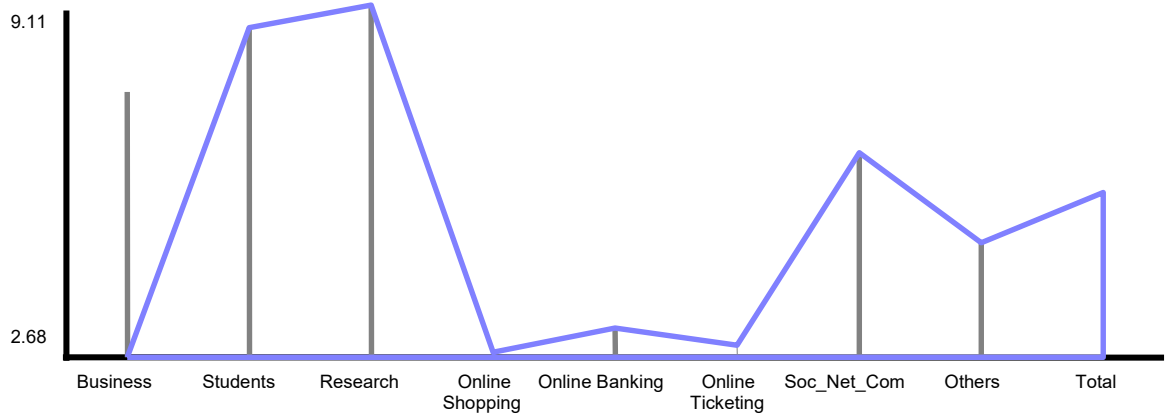

Figure 5: Impacts of IT (Bangladesh)

## 5. CONCLUSIONS

IE Market Research Corp.'s report "1Q10 Bangladesh Mobile Operator Forecast, 2009 – 2014" suggests that Bangladesh will have 72.7 Million mobile phone Subscribers in 2014.[10] To let Bangladeshi people benefit from the internet, it is now obvious to consider wireless broadband technologies because it will not only reduce the hassle of wiring (as most of the area are not facilitated with wired telephony services) but also will save a big amount of money required for polling the cables. In addition to these, rural and geographically diverged areas can easily be covered by the mobile or wireless broadband and it is worth considering. Our study also suggests that the mobile broadband technology holds a great promise in terms of being adopted by the IS users of Bangladesh if the price of the service could be kept in line with their income.